\documentclass[preprint]{jpsj2}

\def\runauthor{
Kenji \textsc{Yonemitsu}
}

\title{
Phase Transition in a One-Dimensional Extended Peierls-Hubbard Model with a Pulse of Oscillating Electric Field: III. Interference Caused by a Double Pulse
}

\author{
Kenji \textsc{Yonemitsu}$^{1,2}$
\thanks{E-mail address: kxy@ims.ac.jp}
}

\inst{
$^1$Institute for Molecular Science, Okazaki 444-8585 \\
$^2$Graduate University for Advanced Studies, Okazaki 444-8585
}

\recdate{\today}

\abst{
In order to study consequences of the differences between the ionic-to-neutral and neutral-to-ionic transitions in the one-dimensional extended Peierls-Hubbard model with alternating potentials for the TTF-CA complex, we introduce a double pulse of oscillating electric field in the time-dependent Schr\"odinger equation and vary the interval between the two pulses as well as their strengths. When the dimerized ionic phase is photoexcited, the interference effect is clearly observed owing to the coherence of charge density and lattice displacements. Namely, the two pulses constructively interfere with each other if the interval is a multiple of the period of the optical lattice vibration, while they destructively interfere if the interval is a half-odd integer times the period, in the processes toward the neutral phase. The interference is strong especially when the pulse is strong and short because the coherence is also strong. Meanwhile, when the neutral phase is photoexcited, the interference effect is almost invisible or weakly observed when the pulse is weak. The photoinduced lattice oscillations are incoherent due to random phases. The strength of the interference caused by a double pulse is a key quantity to distinguish the two transitions and to evaluate the coherence of charge density and lattice displacements. }

\kword{
photoinduced phase transition, neutral-ionic transition, TTF-CA, charge-transfer complex, time-dependent Schr\"odinger equation, dynamics
}

\begin{document}
\sloppy
\maketitle

\section{Introduction}

In paper I \cite{yonemitsu04a}, it is suggested that we can induce the dynamics of charge density and that of lattice displacements simultaneously or separately by choosing the character of the applied electric field. In this sense, the electron-lattice coherence can be controlled. The charge-lattice-coupled dynamics is regarded as coherent if the ionicity and the staggered lattice displacements evolve simultaneously. In this case, the phases of the staggered lattice displacements are almost spatially uniform. In paper II \cite{yonemitsu04b}, we show the difficulty of spontaneously developing charge transfer to the ionic state possibly because the electron-electron interaction is responsible for the charge gap. In the present paper, we show that the degree of coherence is measured by using a double pulse and observing the resultant interference. Thus a double pulse experiment would be a useful probe to characterize the photoinduced dynamics.

In some cases, the degree of coherence may be guessed by comparing experimental data on different space and time scales, like photoreflectance with good space and time resolutions \cite{iwai02} and bulk-sensitive x-ray diffraction for the averaged structure \cite{guerin04} of the TTF-CA complex, although we have to pay attention to different experimental conditions including the energy of the pump light. For instance, in the spin-crossover complex [Fe(2pic)$_3$]Cl$_2$EtOH that shows a two-step thermal transition, although the intermediate-temperature phase is now known to have a symmetry-broken high-spin-low-spin configuration \cite{chernyshov_spin_crossover}, the x-ray diffraction does not so far show a corresponding photoinduced structural change as an averaged structure \cite{collet_spin_crossover}. Such a difference between the thermo- and photo-induced structural changes may be related with different spatial correlations \cite{luty_yonemitsu_spin_crossover}. In the TTF-CA complex, the correlation between the ionicity and the staggered lattice displacements or the spatial correlation of the latter would be important in dynamically controlling the transition in future.

The coherence studied here is different from the coherence of the electronic excitations in quantum optics or from the quantum coherence needed for quantum computing. Even in photoinduced phase transitions, one from Mott insulator to metallic phases, for example, would proceed very fast without structural changes, so that it may have coherence of electronic origin. The ionic-to-neutral transition is, however, necessarily accompanied with structural changes, so that there is basically no coherence of electronic origin. It might have some interference effect when the interval of a double pulse is comparable with the period of the excitonic oscillation, but it would be hardly related with the phase transition. As long as the phase transition is accompanied with structural changes, they determine the transition rate.

Previously, we show a rapid and small-amplitude oscillation of the ionicity in the inset of Fig.~7 in ref.~\citen{miyashita03}. However, only the slow and large-amplitude changes are accompanied with large structural changes. The rapid change corresponds to the excitonic oscillation, whose Fourier spectrum is rather broad [Fig.~8(a) of ref.~\citen{miyashita03}]. Thus, the rapid oscillation cannot keep the electronic coherence on the time scale of the optical lattice vibration [Fig.~7(a) of ref.~\citen{miyashita03}]. In short, the coherence studied here is basically of lattice origin, describing a state where the phases of the staggered lattice displacements are almost uniform in a wide spatial and temporal region inside the ionic domain. 

In paper I for the transition from the ionic phase, the ionicity and the staggered lattice displacements evolve on a common time scale when the pulse is strong and short, while they evolve on different time scales when the pulse is weak and long. If one looks at a certain site, when a strong and short pulse is applied, the lattice displacement oscillates without large friction before and after the neutral-ionic domain wall passes this site and the center of the oscillation is accordingly shifted. We use a double pulse in this paper to show that the interference effect survives for a wide range of pulse intervals when the split pulse is strong and short, while it decays with increasing interval when the split pulse is weak and long. These facts have the same origin, i.e., the coherence is strong when the pulse is strong and short, while it is weak otherwise. In this sense, a double pulse experiment would be suitable for estimating the degree of coherence. In paper II for the transition from the neutral phase, it is shown that the polarization is hardly aligned in the ionic phase. We show here by a double pulse that the interference effect is much weaker than in the transition from the ionic phase. The coherence is then hardly achieved. In this way, the differences between the ionic-to-neutral transition in paper I and the neutral-to-ionic transition in paper II are characterized from the viewpoint of the coherence.

In general, observing the coherence through the interference caused by a double pulse would be useful to study the relation between the dynamics of charge density and that of lattice displacements. If we find a difference between the simulated dynamics and the experimentally observed one through the interference effect, we would be able to construct better models. For instance, three-dimensionality and energy dissipation are missing in this series of papers, and their importance would be evaluated by comparing the numerical results and the experimental data. When we study in future how initially one-dimensional metastable domains finally grow three-dimensionally, the interference effect may give a hint for clarifying the transition processes and for devising a scheme to control them.

\section{Extended Peierls-Hubbard Model with Alternating Potentials}\label{model}

We use a one-dimensional extended Peierls-Hubbard model with alternating potentials at half filling, as in paper I,
\begin{equation}
H =  H_\mathrm{el} + H_\mathrm{lat} \;,\label{g-ham} 
\end{equation}
with
\begin{align}
H_\mathrm{el} = & 
-t_0 \sum _{\sigma,l=1}^{N}
   \left( c^{\dagger }_{l+1,\sigma}c_{l,\sigma}
     + \mathrm{h.c.} \right)  \nonumber \\ 
& +\sum _{l=1}^{N} 
\left[ U n_{l,\uparrow} n_{l,\downarrow} 
+ (-1)^{l} \frac{d}{2} n_{l} \right] \nonumber \\ 
& +\sum _{l:odd}^{N} 
\bar{V}_{l} (n_{l}-2) n_{l+1}
+ \sum _{l:even}^{N} 
\bar{V}_{l} n_{l} (n_{l+1}-2) \;, \label{e-ham} \\
H_\mathrm{lat} =& \sum _{l=1}^{N}
\left[ \frac{k_1}{2}y_{l}^{2}
+\frac{k_2}{4}y_{l}^{4}
+\frac{1}{2}m_{l}\dot{u}_{l}^{2} \right] \;,\label{l-ham} 
\end{align}
where, $ c^{\dagger }_{l,\sigma} $  ($ c_{l,\sigma} $) is the creation (annihilation) operator of a $\pi$-electron with spin $\sigma$ at site $l$, $ n_{l,\sigma} = c^{\dagger}_{l,\sigma} c_{l,\sigma} $, $ n_{l} = n_{l,\uparrow} + n_{l,\downarrow} $, $ u_{l} $ is the dimensionless lattice displacement of the $l$th molecule along the chain from its equidistant position, and $ y_{l} = u_{l+1} - u_{l} $.
The distance between the $l$th and $(l+1)$th molecules is then given by $r_l=r_0 (1+u_{l+1}-u_l)$, where $r_0$ is the averaged distance between the neighboring molecules along the chain. 
The other notations are also the same as in paper I. 
We numerically solve the time-dependent Schr\"odinger equation with the help of the unrestricted Hartree-Fock approximation for the electronic part, and the classical equation of motion for the lattice part, as described in paper I. 
Note that random numbers are added to the initial $ y_{l} $ and $ \dot{u}_l $  values according to the Boltzmann distribution at a fictitious temperature $\it{T}$. 

Photoexcitations are introduced by modifying the transfer integral with the Peierls phase. 
In contrast to papers I and II, a double pulse of electric field is used. 
The first pulse is given by 
\begin{equation}
E(t) = \frac{ E_{\rm ext} }{2} \sin \omega_{\rm ext} t
\;,
\end{equation}
for $ 0 < t < N_{\rm ext} T_{\rm ext} $ with integer $ N_{\rm ext} $, $ T_{\rm ext} = 2 \pi / \omega_{\rm ext} $, and the second pulse by 
\begin{equation}
E(t) = \frac{ E_{\rm ext} }{2} \sin \omega_{\rm ext} ( t - t_{\rm 2nd} )
\;,
\end{equation}
for $ t_{\rm 2nd} < t < t_{\rm 2nd} + N_{\rm ext} T_{\rm ext} $. $ E(t) $ is zero otherwise. If $ t_{\rm 2nd} $ is smaller than $ N_{\rm ext} T_{\rm ext} $, we add up the two terms for $ t_{\rm 2nd} < t < N_{\rm ext} T_{\rm ext} $. Therefore, if $ t_{\rm 2nd} $ is zero, it coincides with the single pulse $ E(t) = E_{\rm ext} \sin \omega_{\rm ext} t $ with amplitude $ E_{\rm ext} $ and frequency $ \omega_{\rm ext} $.

\section{Results and Discussions}\label{results}

We use $N$=100, $ t_0 $=0.17eV, $ U $=1.528eV, $ V $=0.604eV (when we start from the ionic phase) or $ V $=0.600eV (when we start from the neutral phase), $ d $=2.716eV, $ \beta_2 $=8.54eV, $ k_1 $=4.86eV, $ k_2 $=3400eV, and the bare phonon energy $ \omega_{\rm opt} \equiv (1/r_0)(2 k_1/m_r)^{1/2} $=0.0192eV, and impose the periodic boundary condition.
Here $m_r$ is the reduced mass for the donor and acceptor molecules. 
With these parameters, the dimerized ionic phase is stable (metastable) and the neutral phase is metastable (stable) for $ V $=0.604eV ($ V $=0.600eV).
These parameters are the same as in ref.~\citen{miyashita03}, so that the bare phonon energy $ \omega_{\rm opt} $ used here is about five times higher than the optical phonon energy of the TTF-CA complex. 
The ionicity is defined as $ \rho = 1 + (1/N)  \sum_{l=1}^{N} (-1)^{l} \langle n_{l} \rangle $.
The staggered lattice displacement is defined as $ y_{st} = (1/N)  \sum_{l=1}^{N} (-1)^l y_l $.

\subsection{Ionic-to-neutral transition}\label{subsec:I-to-N}

\begin{figure}
\includegraphics[height=12cm]{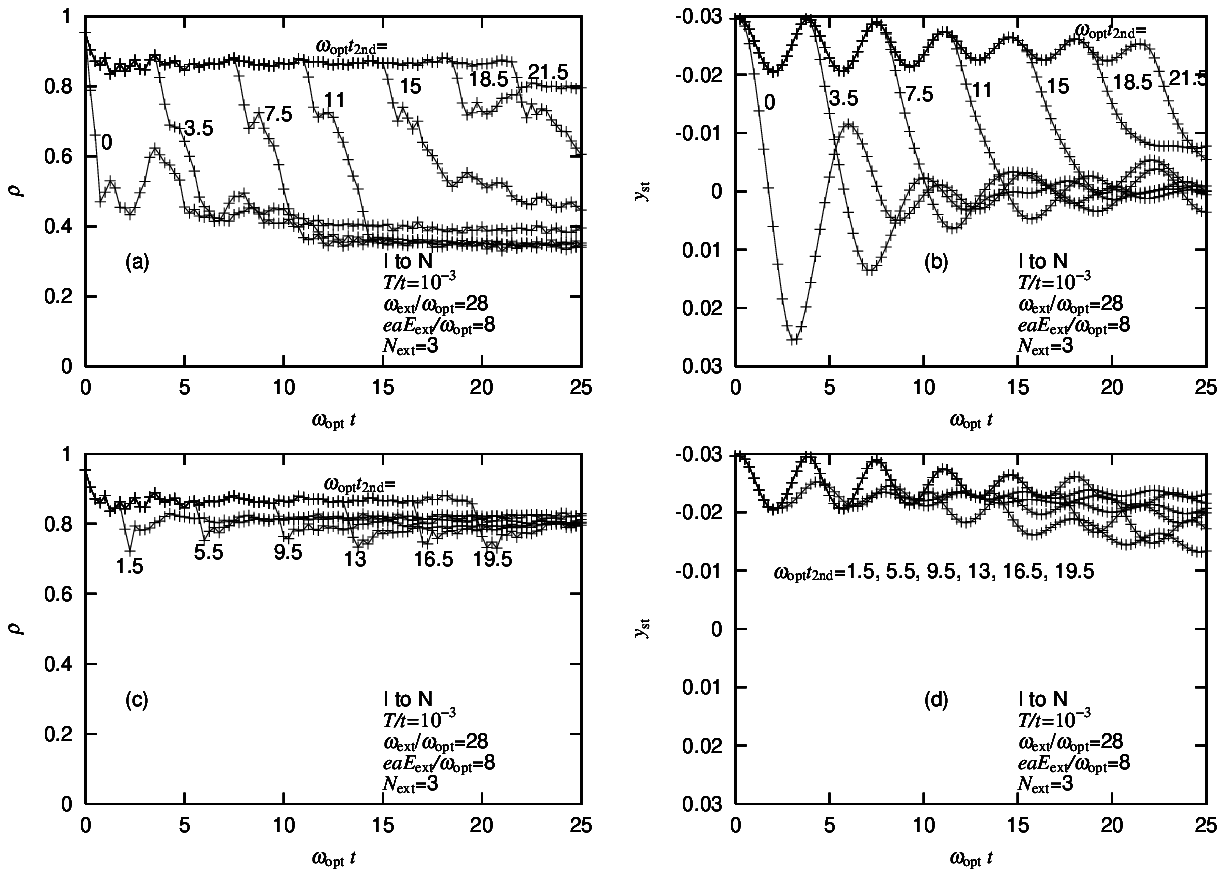}
\caption{Time dependence of (a) the ionicity and (b) the staggered lattice displacement in case the transition is achieved, and that of (c) the ionicity and (d) the staggered lattice displacement in case the transition is not achieved. The interval $ \omega_{\rm opt} t_{\rm 2nd} $ is varied between the two pulses of $ \omega_{\rm ext} / \omega_{\rm opt} $=28, $ eaE_{\rm ext}/\omega_{\rm opt} $=8, and $ N_{\rm ext} $=3. The initial state is ionic at $ T/t $=10$^{-3}$.}
\label{fig:I_N_strong_coherence}
\end{figure}

In this subsection, we adopt the ionic state as the initial condition for the time-dependent Schr\"odinger equation. 
If we use sufficiently weak fields with a given pulse duration, we do not observe any substantial change in the ionicity, irrespective of whether the field is split into two or not, and no matter how long the interval is between the split pulses.
If we use sufficiently strong fields on the other hand, the transition always takes place to the neutral phase.
There is thus a range of the field strengths for which the evolution of the ionicity depends on the interval, as shown below.
By adopting the field strength in this range, we vary the interval between the two pulses to show the time dependence of the ionicity in Figs.~\ref{fig:I_N_strong_coherence}(a) and \ref{fig:I_N_strong_coherence}(c) and that of the staggered lattice displacement in Figs.~\ref{fig:I_N_strong_coherence}(b) and \ref{fig:I_N_strong_coherence}(d). 
As shown in Figs.~\ref{fig:I_N_strong_coherence}(b) and \ref{fig:I_N_strong_coherence}(d), the period of the optical lattice vibration $ T_{\rm opt} $ is slightly less than 4$ / \omega_{\rm opt} $ in the ionic phase.

When the interval between the two pulses $ t_{\rm 2nd} $ is a multiple of $ T_{\rm opt} $, the transition takes place [Figs.~\ref{fig:I_N_strong_coherence}(a) and (b)]. Here, the field is so strong that the single pulse ($ t_{\rm 2nd} $=0) first reverses the polarization [Fig.~\ref{fig:I_N_strong_coherence}(b)] and the reduced ionicity is partially restored after one period [Fig.~\ref{fig:I_N_strong_coherence}(a)]. After two periods, the ionicity is again partially restored but its intensity is weaker than that after one period. The staggered lattice displacement shows decaying oscillation. For $ t_{\rm 2nd} $ being a finite multiple of $ T_{\rm opt} $, the overshoot of the staggered lattice displacement after the second pulse is quickly weakened and soon becomes invisible with increasing $ t_{\rm 2nd} $. Then, the staggered lattice displacement decays almost simultaneously with the ionicity. The speed of the transition slightly and gradually slows down with increasing $ t_{\rm 2nd} $.

On the other hand, when the interval is a half-odd integer times $ T_{\rm opt} $, the transition is suppressed by the destructive interference [Figs.~\ref{fig:I_N_strong_coherence}(c) and (d)]. After the second pulse is applied, the ionicity slightly drops, but it never reaches a value in the neutral phase [Fig.~\ref{fig:I_N_strong_coherence}(c)]. 
In Fig.~\ref{fig:I_N_strong_coherence}(d), one sees that the first and second pulses hit the staggered lattice displacement in the opposite directions. The first pulse decreases the staggered lattice displacement, while the second one increases it. This is why the interference is destructive.

Here we used by chance, for the frequency of the electric field, a multiple of the bare phonon energy. The results are irrespective of whether their ratio is an integer or not. The actual frequency of the optical lattice vibration is different from (higher because of the $ k_2 $ term than) the bare phonon energy, so that the frequency of the electric field is incommensurate with the frequency of the optical lattice vibration. 

\begin{figure}
\includegraphics[height=12cm]{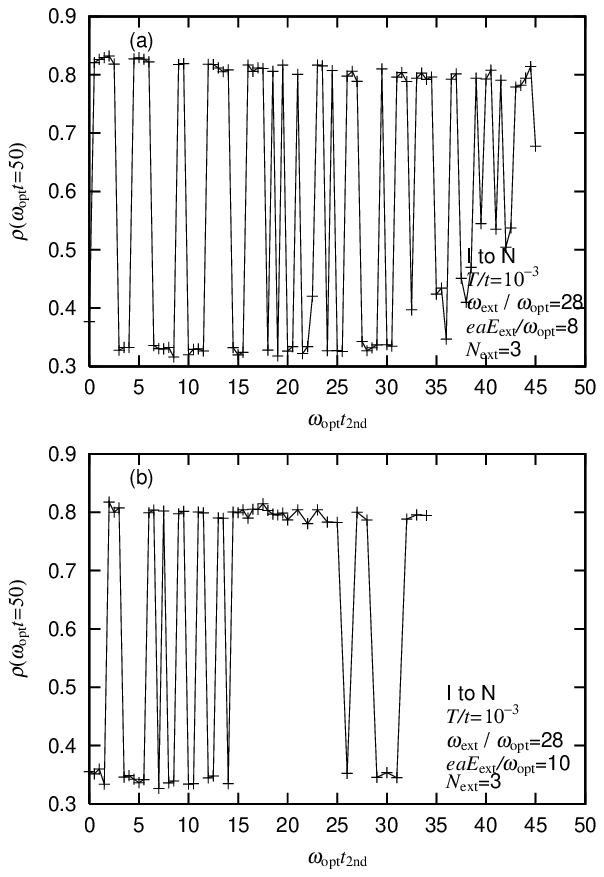}
\caption{Ionicity at $ \omega_{\rm opt} t $=50, as a function of the interval between the two pulses $ \omega_{\rm opt} t_{\rm 2nd} $, (a) for $ eaE_{\rm ext}/\omega_{\rm opt} $=8, and (b) for $ eaE_{\rm ext}/\omega_{\rm opt} $=10. The electric field with $ \omega_{\rm ext} / \omega_{\rm opt} $=28 and $ N_{\rm ext} $=3 is applied to the ionic phase at $ T/t $=10$^{-3}$. The final state is either ionic or neutral.}
\label{fig:I_N_strong_final}
\end{figure}

When the ionic state is photoexcited, the final state is clearly identified as either the ionic state or the neutral state, i.e., no intermediate state is produced, as demonstrated in paper I.
Here we show the ionicity at $ \omega_{\rm opt} t $=50 in Fig.~\ref{fig:I_N_strong_final}, as a function of the interval between the two pulses $ t_{\rm 2nd} $.
The ionicity takes basically two values, as expected, about 0.8 if the final state is ionic and about 0.3 if neutral. 
The high or low value appears almost periodically in Fig.~\ref{fig:I_N_strong_final}(a), and its period is about that of the optical lattice vibration.
For $ \omega_{\rm opt} t_{\rm 2nd} \geq $30 the ionicity in the neutral state becomes significantly larger than 0.3. This is simply because the transition is not completed yet. If we plot the ionicity, e.g., at $ \omega_{\rm opt} t $=100 instead of $ \omega_{\rm opt} t $=50, it is again about 0.3.
For short pulses such as we presently adopt ($ N_{\rm ext} $=3), the range of the field strengths for which the interference is clearly observed is wide.
In Fig.~\ref{fig:I_N_strong_final}(b), we use a stronger pulse to show the similar result.
Although the final ionicity is not as periodic as in Fig.~\ref{fig:I_N_strong_final}(a), the interference is clearly seen even for $ \omega_{\rm opt} t_{\rm 2nd} \simeq $30.
The deviation from the periodicity is due to the fact that the staggered lattice displacement is not a pure sinusoidal function with a single frequency (not shown). Because other frequencies are substantially involved for the stronger pulse, the amplitude of the oscillation of the staggered lattice displacement changes with time after the first pulse. When the amplitude is small, the interference effect is suppressed as in Fig.~\ref{fig:I_N_strong_final}(b) for $ \omega_{\rm opt} t_{\rm 2nd} \simeq $20. We have performed extensive numerical calculations with different strengths, durations and intervals of pulses. The periodicity as a function of the interval is lost as shown here in an accidental manner. The oscillation of the staggered lattice displacement contains different frequency components around the main one, which seem to have stochastic distribution depending on the random numbers added to the initial lattice variables. 

\begin{figure}
\includegraphics[height=12cm]{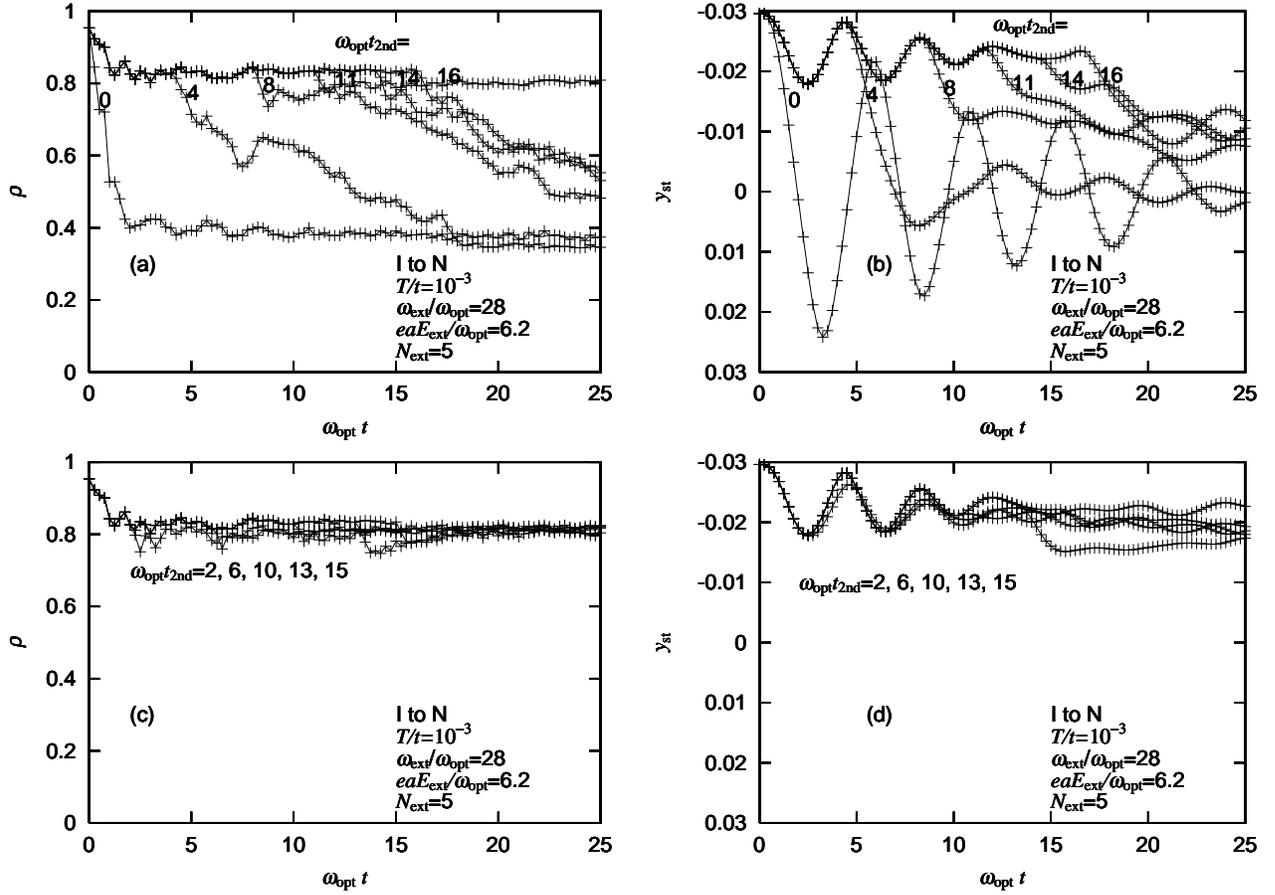}
\caption{Time dependence of (a) the ionicity and (b) the staggered lattice displacement in case the transition is achieved, and that of (c) the ionicity and (d) the staggered lattice displacement in case the transition is not achieved. The interval $ \omega_{\rm opt} t_{\rm 2nd} $ is varied between the two pulses of $ \omega_{\rm ext} / \omega_{\rm opt} $=28, $ eaE_{\rm ext}/\omega_{\rm opt} $=6.2, and $ N_{\rm ext} $=5, which are weaker and longer than in Fig.~\ref{fig:I_N_strong_coherence}. The initial state is ionic at $ T/t $=10$^{-3}$. The coherence is slightly weaker than in Fig.~\ref{fig:I_N_strong_coherence}.}
\label{fig:I_N_weak_coherence}
\end{figure}

Now we use a weaker and longer pulse to show how the interference effect is weakened. 
The range of the field strengths for which the final state depends on the interval is substantially narrower than before.
The time dependence of the ionicity is shown in Figs.~\ref{fig:I_N_weak_coherence}(a) and \ref{fig:I_N_weak_coherence}(c), and that of the staggered lattice displacement in Figs.~\ref{fig:I_N_weak_coherence}(b) and \ref{fig:I_N_weak_coherence}(d).
After the first pulse, the oscillation of the staggered lattice displacement decays faster [Fig.~\ref{fig:I_N_weak_coherence}(b)].
Even when the second pulse is applied only $ T_{\rm opt} $ after the first pulse, the staggered lattice displacement shows only a weak overshoot and its oscillation in the neutral phase is incoherent.
When the second pulse is applied $ 2 T_{\rm opt} $ after the first pulse, the transition is not smooth any more and it takes a much longer time for the staggered lattice displacement to disappear.
The evolution of the ionicity after the second pulse is also rather irregular and much slower [Fig.~\ref{fig:I_N_weak_coherence}(a)] than before.
Nevertheless, the coherence is still clearly observed.
When the interval is a half-odd integer times $ T_{\rm opt} $, the second pulse little affects the ionicity [Fig.~\ref{fig:I_N_weak_coherence}(c)] and the staggered lattice displacement [Fig.~\ref{fig:I_N_weak_coherence}(d)].

\begin{figure}
\includegraphics[height=12cm]{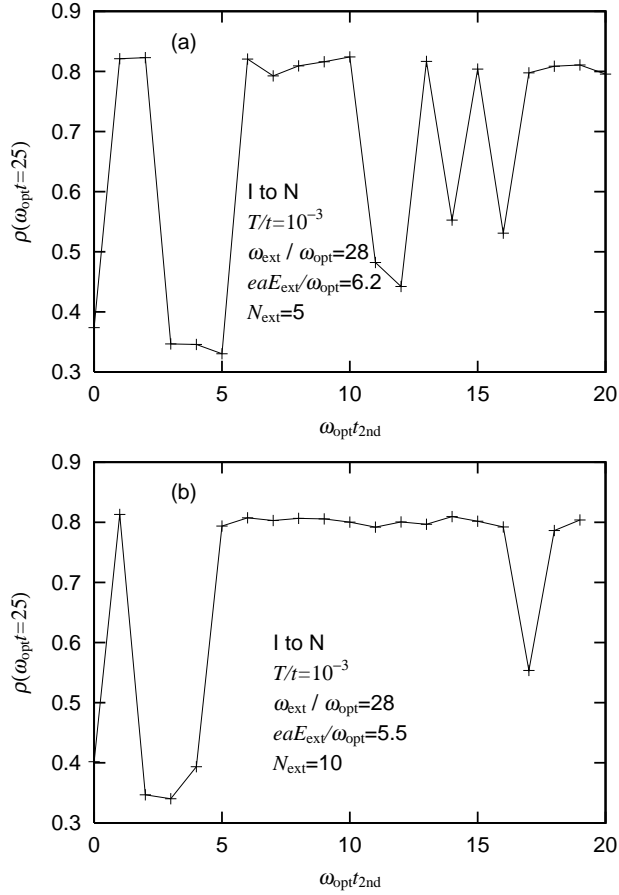}
\caption{Ionicity at $ \omega_{\rm opt} t $=25, as a function of the interval between the two pulses $ \omega_{\rm opt} t_{\rm 2nd} $, (a) for $ eaE_{\rm ext}/\omega_{\rm opt} $=6.2 and $ N_{\rm ext} $=5, and (b) for $ eaE_{\rm ext}/\omega_{\rm opt} $=5.5 and $ N_{\rm ext} $=10. The electric field with $ \omega_{\rm ext} / \omega_{\rm opt} $=28 is applied to the ionic phase at $ T/t $=10$^{-3}$. The final state is still either ionic or neutral.}
\label{fig:I_N_weak_final}
\end{figure}

With the same strength and duration of the pulse as in Fig.~\ref{fig:I_N_weak_coherence}, we show the ionicity at $ \omega_{\rm opt} t $=25 in Fig.~\ref{fig:I_N_weak_final}(a), as a function of the interval between the two pulses $ t_{\rm 2nd} $. Again, the final ionicity takes basically two values, about 0.8 and 0.3. The rather large values around 0.5 in the neutral state for $ \omega_{\rm opt} t_{\rm 2nd} \geq $10 is due to the fact that the transition is not completed yet. If we plot the really final ionicity instead, they become about 0.3. In any case, the final ionicity largely deviates from the periodic function. This is again due to the fact that the staggered lattice displacement is not purely sinusoidal with a single frequency [Fig.~\ref{fig:I_N_weak_coherence}(b)]. Because other frequencies are substantially involved for the longer pulse (even if it is weak), the amplitude of the oscillation of the staggered lattice displacement substantially changes with time after the first pulse, as clearly seen in Fig.~\ref{fig:I_N_weak_coherence}(b). As the pulse becomes longer (and weaker to maintain the number of absorbed photons), the final ionicity more strongly deviates from the periodic function [Fig.~\ref{fig:I_N_weak_final}(b)]. In addition, the range of the field strengths for which the final state depends on the interval becomes narrow.

\subsection{Neutral-to-ionic transition}\label{subsec:N-to-I}

\begin{figure}
\includegraphics[height=12cm]{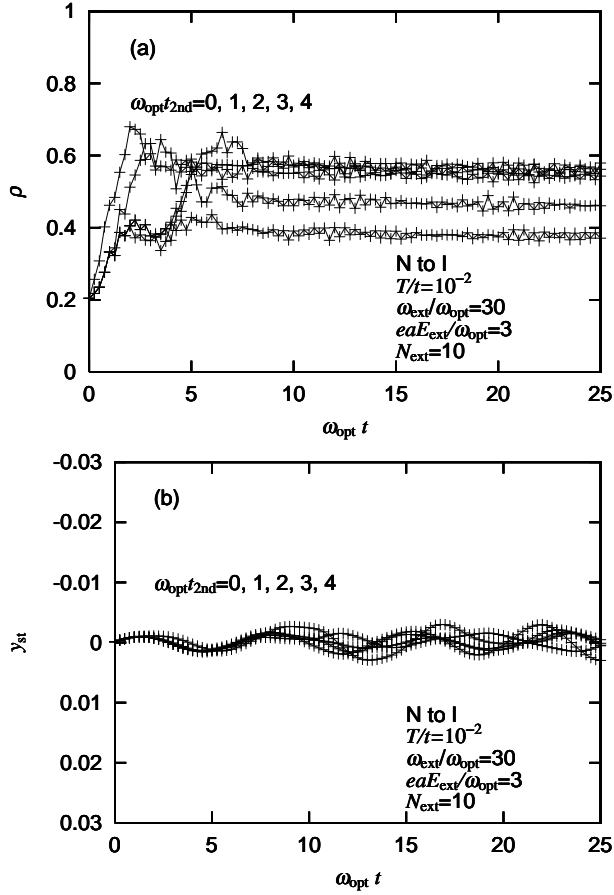}
\caption{Time dependence of (a) the ionicity and (b) the staggered lattice displacement. The interval $ \omega_{\rm opt} t_{\rm 2nd} $ is varied between the two pulses of $ \omega_{\rm ext} / \omega_{\rm opt} $=30, $ eaE_{\rm ext}/\omega_{\rm opt} $=3, and $ N_{\rm ext} $=10. The initial state is neutral at $ T/t $=10$^{-2}$. The coherence is almost completely lost.}
\label{fig:N_I_weak_incoherence}
\end{figure}

\begin{figure}
\includegraphics[height=12cm]{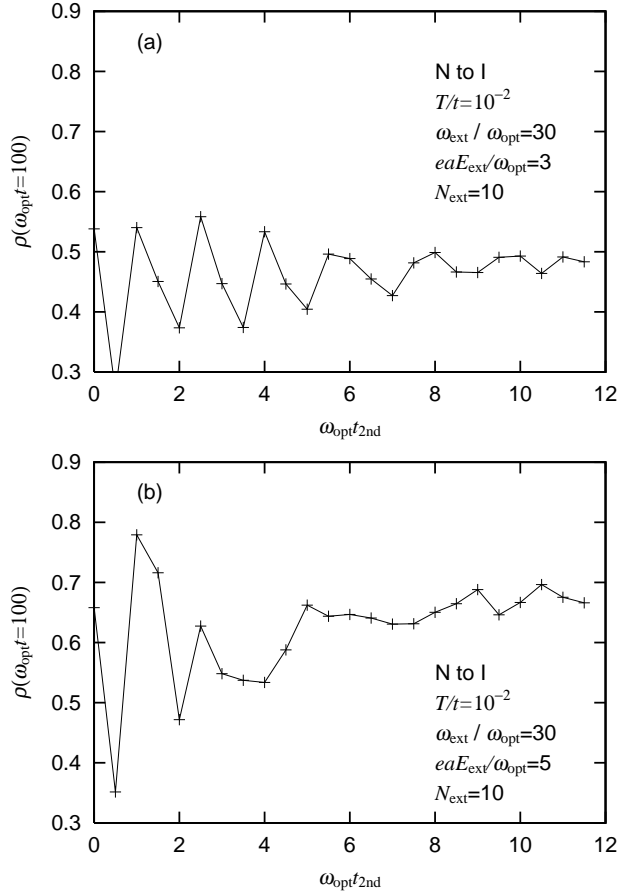}
\caption{Ionicity at $ \omega_{\rm opt} t $=100, as a function of the interval between the two pulses $ \omega_{\rm opt} t_{\rm 2nd} $, (a) for $ eaE_{\rm ext}/\omega_{\rm opt} $=3, and (b) for $ eaE_{\rm ext}/\omega_{\rm opt} $=5. The electric field with $ \omega_{\rm ext} / \omega_{\rm opt} $=30 and $ N_{\rm ext} $=10 is applied to the neutral phase at $ T/t $=10$^{-2}$. The final state changes continuously with $ eaE_{\rm ext}/\omega_{\rm opt} $.}
\label{fig:N_I_weak_final}
\end{figure}

In this subsection, the initial state is neutral and its lattice displacements are distributed around zero in equilibrium. In contrast to the previous subsection, the clear interference effect is not seen with any field strength. The final ionicity is much more insensitive to the interval between the two pulses. Nonetheless, a small-amplitude oscillation is found as a function of the interval, as shown later. We show the time dependence of the ionicity in Fig.~\ref{fig:N_I_weak_incoherence}(a) and that of the staggered lattice displacement in Fig.~\ref{fig:N_I_weak_incoherence}(b) with different intervals. The first pulse is applied for $ 0 \leq \omega_{\rm opt} t \leq N_{\rm ext} \omega_{\rm opt} T_{\rm ext} $=2.1. Then, for $ \omega_{\rm opt} t_{\rm 2nd} \leq $2, the two pulses directly interfere with each other to show the superficial dependence on $ \omega_{\rm opt} t_{\rm 2nd} $. Otherwise, the ionicity is insensitive to $ \omega_{\rm opt} t_{\rm 2nd} $ [Fig.~\ref{fig:N_I_weak_incoherence}(a)]. To assign $ \omega_{\rm opt} t_{\rm 2nd} $ to each curve, refer to Fig.~\ref{fig:N_I_weak_final}(a) later. The reason for the insensitivity is rather obvious when one sees the evolution of the staggered lattice displacement [Fig.~\ref{fig:N_I_weak_incoherence}(b)]. 
Since the initial lattice displacements are distributed around zero, they start to oscillate in a random manner. Some displacements contribute to positive $ y_{\rm st} $, others to negative $ y_{\rm st} $. Due to the cancellation, the magnitude of $ y_{\rm st} $ is small. The second pulse also affects the displacements in a very similar manner. Then, the interference effect is hardly seen when the staggered lattice displacements are spatially averaged.

When the neutral state is photoexcited, the final state continuously changes with the strength and duration of the pulse, as demonstrated in paper II. 
The ionicity at $ \omega_{\rm opt} t $=100 is shown in Fig.~\ref{fig:N_I_weak_final}, as a function of the interval between the two pulses $ t_{\rm 2nd} $.
The final ionicity continuously changes with $ t_{\rm 2nd} $ and also with the field strength, as expected. 
With the stronger field, it is larger on average [Fig.~\ref{fig:N_I_weak_final}(b) compared with Fig.~\ref{fig:N_I_weak_final}(a)].
With the weaker field, the final ionicity appears to oscillate [Fig.~\ref{fig:N_I_weak_final}(a)], although the amplitude of the oscillation is much smaller than that in the previous subsection. 
An interference effect is experimentally observed in the photoinduced dynamics from the neutral phase \cite{okamoto_private}, though it is not compared yet with the interference effect expected in the photoinduced dynamics from the ionic phase. 
The numerically observed period is about a half of $ T_{\rm opt} $ in contrast to $ T_{\rm opt} $ in the previous subsection. 

\begin{figure}
\includegraphics[height=12cm]{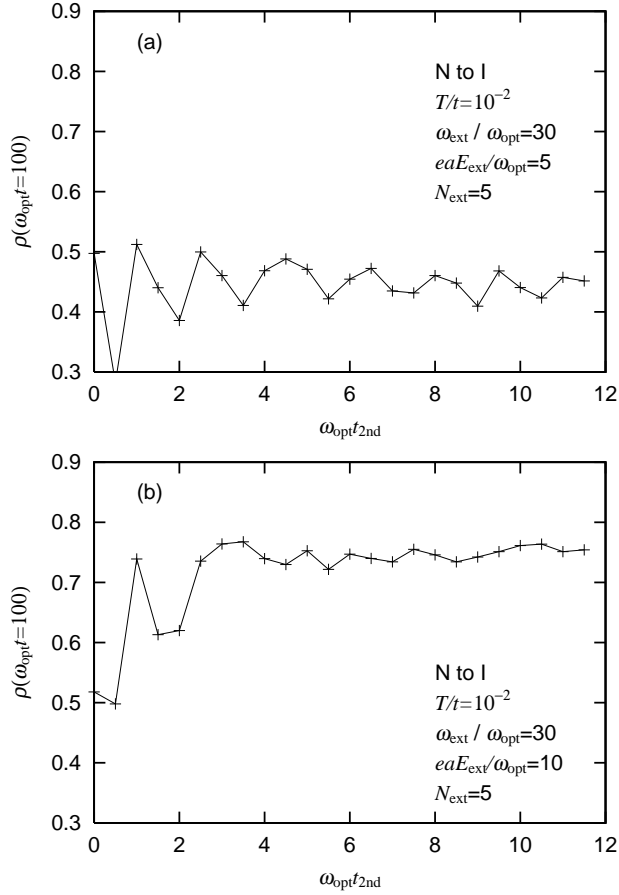}
\caption{ Ionicity at $ \omega_{\rm opt} t $=100, as a function of the interval between the two pulses $ \omega_{\rm opt} t_{\rm 2nd} $, (a) for $ eaE_{\rm ext}/\omega_{\rm opt} $=5, and (b) for $ eaE_{\rm ext}/\omega_{\rm opt} $=10. The electric field with $ \omega_{\rm ext} / \omega_{\rm opt} $=30 and $ N_{\rm ext} $=5, which is stronger and shorter than in Fig.~\ref{fig:N_I_weak_final}, is applied to the neutral phase at $ T/t $=10$^{-2}$.}
\label{fig:N_I_strong_final}
\end{figure}

To demonstrate that this small-amplitude oscillation is generally found with weak fields, we use shorter pulses to show the ionicity at $ \omega_{\rm opt} t $=100 in Fig.~\ref{fig:N_I_strong_final}, as a function of the interval. The results are similar. Namely, the final ionicity continuously changes with $ t_{\rm 2nd} $ and with the field strength. The final ionicity is larger for the stronger field. Again with the weaker field, the final ionicity oscillates [Fig.~\ref{fig:N_I_strong_final}(a)]. The amplitude and period of this oscillation is similar to those in Fig.~\ref{fig:N_I_weak_final}(a).

\begin{figure}
\includegraphics[height=6 cm]{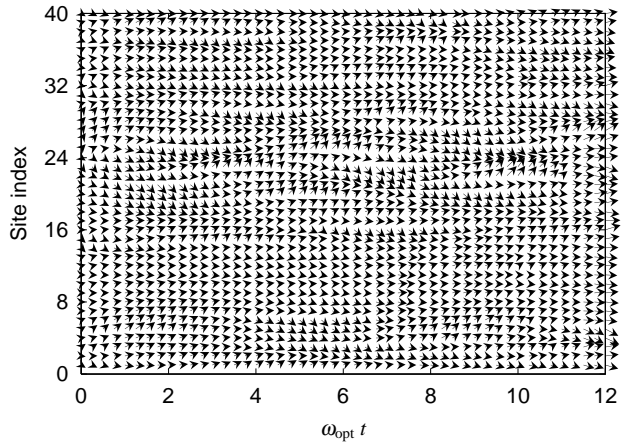}
\caption{Correlation between the staggered lattice displacement $ y_{st \; l} $ (the vertical component of the bar) and the ionicity $ \rho_l $ (the horizontal component of the bar), as a function of the site index $ l $ and the elapsing time $ t $ multiplied by $ \omega_{\rm opt} $. The electric field with $ eaE_{\rm ext}/\omega_{\rm opt} $=5, $ N_{\rm ext} $=5 and of frequency $ \omega_{\rm ext} / \omega_{\rm opt} $=30 is split into two with an interval of $ \omega_{\rm opt} t_{\rm 2nd} $=11 and is applied to the neutral phase at $ T/t $=10$^{-2}$.}
\label{fig:N_I_space_time_dependence}
\end{figure}

The origin of the small-amplitude oscillation of the spatially averaged, staggered lattice displacement is rather obvious when one locally sees the staggered lattice displacement. The space and time evolution of the ionicity and the staggered lattice displacement is shown in Fig.~\ref{fig:N_I_space_time_dependence}. The horizontal component of each bar represents the local ionicity $ \rho_l $ defined as $ \rho_l = 1 + (-1)^l ( - \langle n_{l-1} \rangle + 2 \langle n_l \rangle - \langle n_{l+1} \rangle )/4 $. The vertical component gives the local staggered lattice displacement $ y_{st \; l} $ defined as $ y_{st \; l} = (-1)^l ( - y_{l-1} + 2 y_l - y_{l+1} )/4 $. The bars are shown on all sites in the first 40 sites. The oscillating electric field is applied for $ 0 \leq \omega_{\rm opt} t \leq 1.05 $ and $ 11 \leq \omega_{\rm opt} t \leq 12.05 $.
It is clearly seen that the first pulse induces oscillations of the local staggered lattice displacements and that their phases are randomly distributed over the system. Because nearby oscillations with different phases compete with each other, each oscillation is not regular if one sees it very carefully. Nevertheless, the local oscillations persist in this time scale. Therefore, if the second pulse is applied a multiple of $ T_{\rm opt} $ after the first pulse, as shown here, it enhances the local oscillations so that the (spatially averaged) final ionicity becomes a little bit larger [compare $ \omega_{\rm opt} t_{\rm 2nd} $=11 in Fig.~\ref{fig:N_I_strong_final}(a) with the nearby points at $ \omega_{\rm opt} t_{\rm 2nd} $=10.5 and 11.5.]. 

It should be noted that, if the second pulse is applied a half-odd integer times $ T_{\rm opt} $ after the first pulse, the final ionicity becomes a little bit larger again owing to constructive interference, in contrast to the destructive interference when the ionic state is photoexcited. One can imagine that a pendulum starts to oscillate from the bottom of a potential by a pulse of external force. When the second pulse is applied a half-odd integer times $ T_{\rm opt} $ after the first pulse, the pendulum is at the bottom of the potential, but its velocity is opposite to the initial one. Note that the electric field induces charge transfer in both directions. The direction to which the pendulum is hit makes no sense. Thus, the two pulses constructively interfere with each other also when the interval is a half-odd integer times $ T_{\rm opt} $. As a consequence, the small-amplitude oscillation appears in the (spatially averaged) final ionicity and its period is about a half of $ T_{\rm opt} $ [Figs.~\ref{fig:N_I_weak_final}(a) and \ref{fig:N_I_strong_final}(a)].
If the electric field is substantially larger, however, the cancellation among different phases of local oscillations is more effective, so that the small-amplitude oscillation becomes invisible [Figs.~\ref{fig:N_I_weak_final}(b) and \ref{fig:N_I_strong_final}(b)].

\section{Conclusions}\label{conc}

In papers I and II, we have shown the qualitative differences between the ionic-to-neutral \cite{yonemitsu04a} and neutral-to-ionic \cite{yonemitsu04b} transitions and the field-strength-dependent relative dynamics of charge density and lattice displacements \cite{yonemitsu04a} in the one-dimensional extended Peierls-Hubbard model with alternating potentials. To study their consequences and to help us devise a scheme to control the transition dynamics, we here use a double pulse of oscillating electric field with different intervals in solving the time-dependent Schr\"odinger equation for the mean-field electronic wave function. 

When the photoexcitation induces the transition from the dimerized ionic phase with ferroelectrically ordered polarizations to the neutral phase with almost equidistant molecules, the interference effect is clearly observed due to the coherence of the ionicity and the staggered lattice displacements. The transition is promoted when the interval between the two pulses is a multiple of the period of the optical lattice vibration, while it is suppressed when the interval is a half-odd integer times the period. The lattice vibrations around the dimerized structure induced by the first and second pulses are in phase or out of phase, respectively. In paper I, the infrared light nearly resonating with the optical lattice vibrations is shown to induce the transition as well as the light for creating excitons. The coupling between charge density and lattice displacements is strong enough to transfer charge by lattice oscillation. In this paper, the relative phase between the lattice vibrations and the oscillating electric field is shown important for the photoinduced transition. 

In paper I, the correlation between the dynamics of the charge density and that of the lattice displacements is shown to depend on the strength of the pulse. When the pulse is strong and short, the charge transfer takes place on the same time scale with the disappearance of dimerization. When the pulse is weak and long, the dimerization-induced polarization is disordered much before the charge transfer. In the present calculations with a double pulse, the coherence of the ionicity and the staggered lattice displacements is also shown to depend on the strength of the pulse. When the pulse is strong and short, the coherence is so strong that the interference effect is observed even for the intervals much longer than the period of the optical lattice vibration. When the pulse is weak and long, the coherence is weakened, and then the interference effect fades away with increasing interval. These findings have the same origin. Therefore, the double-pulse experiment would be a good tool to investigate the coherence of the ionicity and the staggered lattice displacements. 

When the photoexcitation induces the transition from the neutral phase with thermally disordered lattice displacements to the ionic phase, a very weak interference effect is observed, although the staggered displacements are spatially incoherent. The induced ionic state is stabilized by staggered lattice displacements, but they need not be spatially coherent in one dimension. The induced polarizations are thus disordered. The final ionicity depends weakly on the interval between the two pulses. Nevertheless, when the field is not so strong, the weak interference effect is certainly observed. The final ionicity oscillates, as a function of the interval, with small amplitude and twice as frequently as observed in the ionic-to-neutral transition. To realize the constructive interference, the second pulse must be applied when the lattice displacements have the maximum velocity. It is hard to observe the interference if the staggered lattice displacements are spatially averaged, but it is relatively easy if they are measured locally. As the pulse is strengthened, different phases of nearby oscillations compete more strongly with each other and suppress the interference effect. 

When quantitatively comparing the numerical results with the experimental observation, three-dimensionality and energy dissipation are not negligible. The double pulse experiment may be very useful in that it allows comparisons from the viewpoint of coherence. Because the information on the coherence of the ionicity and the staggered lattice displacements is obtained by the double pulse experiment, the relevance of the isolated one-dimensional model system can be discussed in a straightforward manner. The relevance may depend on the time scale of observation.

\section*{Acknowledgement}

The author is grateful to H. Okamoto for showing his data prior to publication and for enlightening discussions. 
This work was supported by Grants-in-Aid for Scientific Research (C) (No. 15540354), for Scientific Research on Priority Area ``Molecular Conductors'' (No. 15073224), for Creative Scientific Research (No. 15GS0216), and NAREGI Nanoscience Project from the Ministry of Education, Culture, Sports, Science and Technology, Japan.



\end{document}